\documentclass[epj]{svjour}

\usepackage{amsmath}
\usepackage{amssymb}
\usepackage{graphicx}
\usepackage{hyperref}
\usepackage[all]{hypcap}
\usepackage{subfigure}
\begin{document}

\title{Properties of the United States Code Citation Network}

\author{Michael J. Bommarito II \inst{1} \inst{2} \inst{3} \and Daniel Martin Katz \inst{1} \inst{3} \inst{4}}
\institute{Department of Political Science, University of Michigan, Ann Arbor \and Department of Mathematics, University of Michigan, Ann Arbor \and Center for the Study of Complex Systems, University of Michigan, Ann Arbor \and University of Michigan Law School}
\date{\today}

\abstract{The \textit{United States Code} (Code) is an important source of Federal law that is produced by the interactions of many heterogeneous actors in a complex, dynamic space.  The Code can be represented as the union of a hierarchical network and a citation network over the vertices representing the language of the Code.  In this paper, we investigate the properties of the Code's citation network by examining the directed degree distributions of the network.  We find that the power-law model is a plausible fit for the outdegree distribution but not for the indegree distribution.  In order to better understand this result, we construct a model with the assumption that the probability of citation is a per-word rate.  We calculate the adjusted degree of each vertex under this model and study the directed adjusted degree distributions.  These adjusted degree distributions indicate that both the adjusted indegree and outdegree distributions seems to follow a log-normal form, not a power-law form.  Our findings indicate that the power-law is not generally applicable to degree distributions within the United States Code but that the distribution of degree per word is well-described by a log-normal model.}


\maketitle

\section{Introduction}
Many real systems are characterized by the interactions of heterogeneous actors operating in a complex, dynamic space.  One important example of such a system is the legislative system of the United States.  The actors in this system include members of Congress and the Federal Government, as well as the wide range of individuals, organizations, and even countries who act to further their own policy goals.  Furthermore, the space of all possible policy configurations is combinatorially complex and nuanced beyond formalization.  One feasible approach to studying this legislative system is to examine a representation of output - the \textit{United States Code} (Code).

The Code is a 22 million word compilation of legislation and treaties that have been ratified by Congress.\footnote{This compilation is performed through a process known as codification, which is carried out by the \textit{Office of the Law Revision Counsel} (LRC), an organization within the U.S. House of Representatives.  2 U.S.C. \S 285- \S285g outlines the purpose, policy and functions of the Office of Law Revision Counsel.  The LRC's goal in this codification process is to transform the incremental and chronological \textit{Statutes at Large} into the Code, a current snapshot of the law that is organized into hierarchical categories.  The Code is only \textit{prima facie} evidence of Federal law.  In the event of a discrepancy, the \textit{Statutes at Large} are the final authority.  Furthermore, additional sources such as the \textit{Code of Federal Regulations} contains clarifications issued by other Federal agencies or bodies.}  This compilation can be mathematically represented as the union of a hierarchical network and a citation network over the vertices representing the language of the Code (\cite{Bommarito2010}).  The purpose of this paper is to describe the properties of this citation network and explore possible explanatory dynamics.

In the remainder of the paper, we first describe the dataset and provide summary statistics of the Code's citation network in Section \ref{sec:data}.  In Section \ref{sec:degree}, we assess whether the power-law is a plausible model for the directed degree distributions.  In Section \ref{sec:adjdegree}, we consider an alternative hypothesis that the governing dynamic is instead ``degree per word.''  We then test both a power-law and log-normal model for this adjusted ``degree per word'' data.  Finally, we conclude with a summary of results in Section \ref{sec:conclusion}.  

\section{Data}
\label{sec:data}
In order to construct a mathematical representation $\mathcal{G} = (V,E)$ of the Code we have obtained an XML snapshot of the Code from March 2010.  This data was provided by the Legal Information Institute at the Cornell University Law School (\cite{Cornell2009}).  Citations are explicitly coded within these XML documents at the section level.  Sections in the Code are similar in purpose to sections within articles, such as this ``Data'' section.  Within the Code, they are unique among possible units of analysis in that they are guaranteed to contain complete grammatically parsable legal text.  Therefore, they make a natural unit of analysis for evaluating linguistic dynamics (\cite{Bommarito2010}).

The Code has 59,988 sections.  However, since some sections are isolates, the largest weakly connected component of the section citation network has only $|V| = 34,674$ vertices.  The total number of edges in the network is $|E| = 85,921$.  These figures imply a density of $7.15 x 10^{-5}$, indicating that the network is relatively sparse.

\section{Degree}
\label{sec:degree}
When presented with a network, the first step is often to study its degree distribution.  Characterizations of the degree distributions are both useful in explaining macroscopic properties of the network and straightforward to calculate (\cite{Newman2003}).

At first glance, the indegree and outdegree distributions appear highly skewed, with skewness values of 15.38 and 6.07 respectively.  When such skewed degree distributions are encountered, power-law distributions are often present.  To test for the existence of the power-law phenomenon in this network, we apply the methods of Clauset et al. (\cite{Clauset2009}) and assess the scaling parameter $\alpha$ and $p$-value of each fit.  These methods are designed to determine whether the data follow the form
\begin{align}
	p(\delta) \propto& \delta^{-\alpha}
\end{align}
where $\delta$ is degree and $\alpha$ is the scaling exponent.  $p$-values are obtained by the software provided in \cite{Clauset2009}. 

\begin{figure}[htb]
	\centering
	\includegraphics[width=7.5cm]{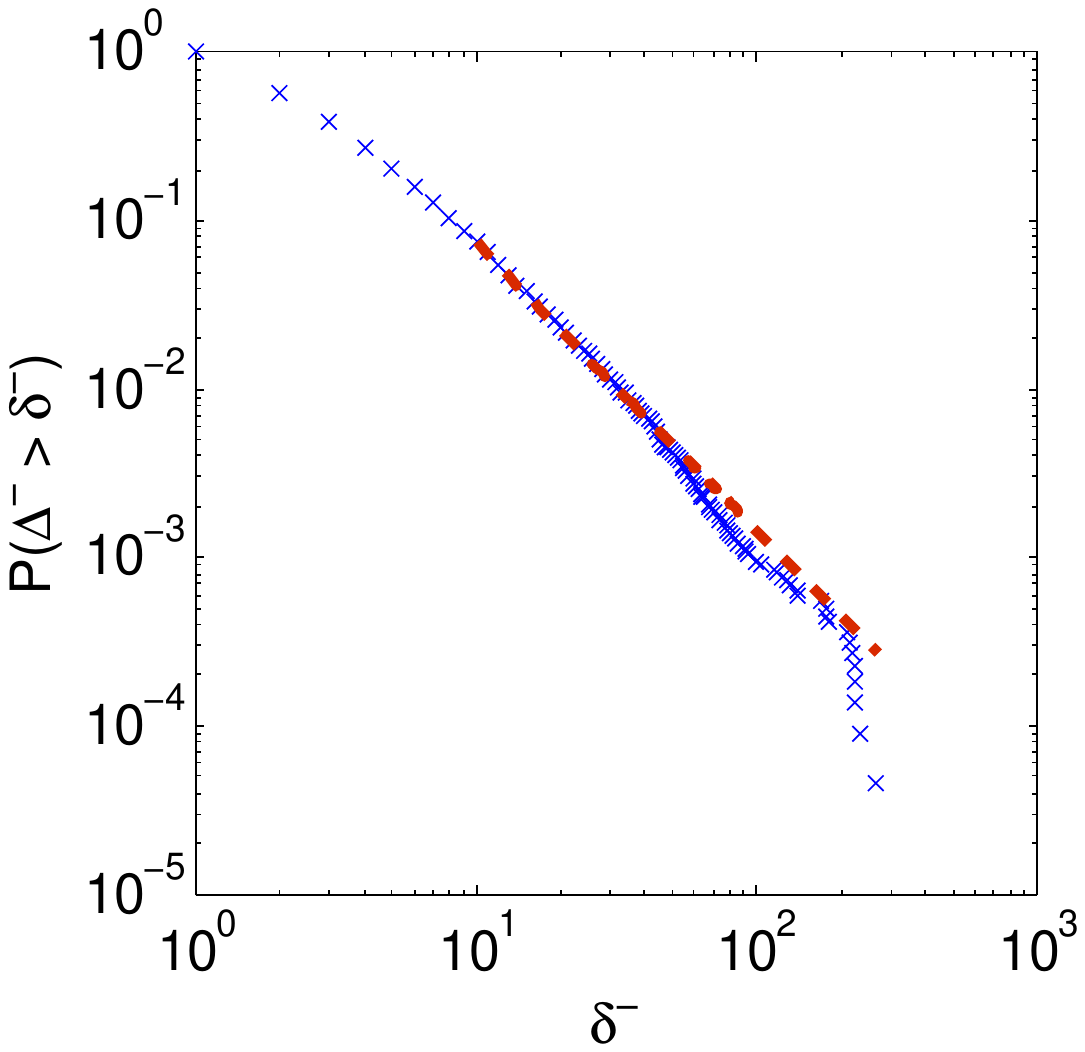}
	\caption{Log-Log Indegree Distribution with Power-Law Fit.}
	\label{fig:indegree}
	
	\includegraphics[width=7.5cm]{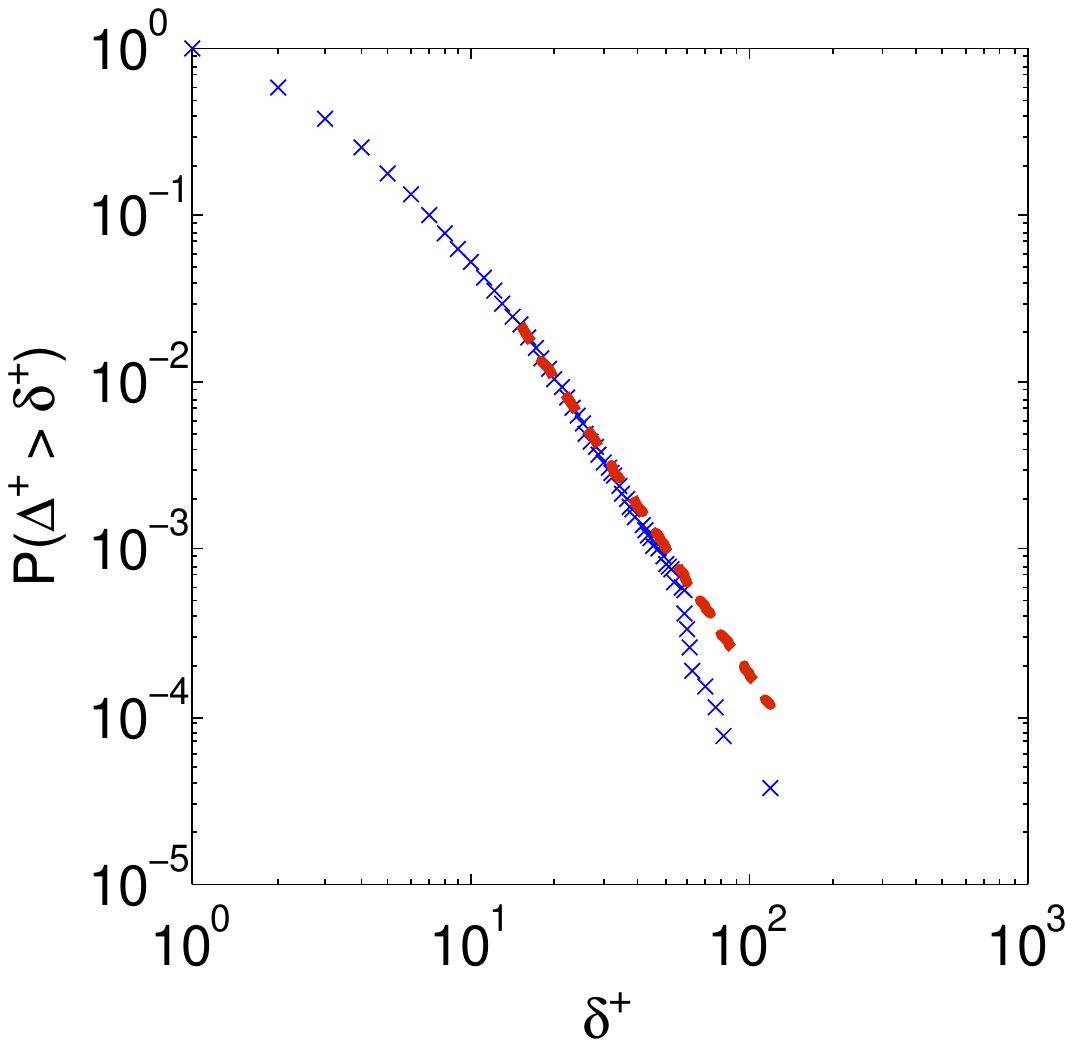}
	\caption{Log-Log Outdegree Distribution with Power-Law Fit.}
	\label{fig:outdegree}
\end{figure}

\begin{table}[htp]
	\centering
	\begin{tabular}{|c|c|c|}
		\hline
		\textbf{Section} & $\delta^{-}$ & $\delta^{+}$\\\hline
		26 U.S.C. 501 & 261 & 42\\\hline
		5 U.S.C. 552 & 231 & 4\\\hline 
		6 U.S.C. 1 & 227 & 1\\\hline 
		6 U.S.C. 3 & 226 & 4\\\hline
		16 U.S.C. 2 & 223 & 2\\\hline
	\end{tabular}
	\caption{Five Highest Indegree Sections}
	\label{table:indegree}

	\begin{tabular}{|c|c|c|}
		\hline
		\textbf{Section} & $\delta^{-}$ & $\delta^{+}$\\\hline
		18 U.S.C. 2516 & 4 & 119\\\hline
		18 U.S.C. 1956 & 28 & 80\\\hline
		42 U.S.C. 1396a & 63 & 76\\\hline
		15 U.S.C. 78c & 73 & 68\\\hline
		18 U.S.C. 1841 & 0 & 64\\\hline
	\end{tabular}
	\caption{Five Highest Outdegree Sections}
	\label{table:outdegree}
\end{table}

Figure \ref{fig:indegree} shows the log-log indegree distribution and its power-law fit as calculated by \cite{Clauset2009}.  $\alpha$ and $x_{min}$ are $2.68$ and $10$ respectively.  Though $\alpha$ falls between 2 and 3, the $p$-value for this model is $0.084$ and the region of fit spans only one order of magnitude.  It thus seems fairly implausible that the data come from a power-law process.  This deviation is most likely due to the steep shift in slope at the tail of the data.  Table \ref{table:indegree} lists the sections in this tail.  These sections corresponds to some of the most well-known and broadly applicable portions of Federal statutory law.  For example, the most cited section, 26 U.S.C. \S 501, contains \S 501(c)(3), the clause defining the conditions for tax-exempt organizations.  Provisions throughout the Code rely on this definition. 

Figure \ref{fig:outdegree} plots the log-log outdegree distribution and its power-law fit.  $\alpha$ and $x_{min}$ are $3.5$ and $15$ respectively.  In this case, the $p$-value is 0.58 and it thus seems plausible that the data is generated by a power-law.  However, the region of fit is only one order of magnitude and the scaling parameter $\alpha$ is greater than $3$.  This implies that these data are an extreme case of the power-law.  The value of $\alpha = 3.5$ is greater than all but 3 of the 17 identified power laws in \cite{Clauset2009}.

Just as in the indegree distribution, much of the deviation is produced in the steep slope at the end of the tail.  Table \ref{table:outdegree} lists the sections that contribute most to this tail.  Many of these sections deal with investigating criminal conduct and thus list the activities which allow for the involvement of Federal law enforcement.  Though the degree distributions above exhibit significant skewness, these results indicate that degree distributions of sections cannot simply be modeled as power-law processes.

\section{Adjusted Degree}
\label{sec:adjdegree}
In Section \ref{sec:degree} above, we model the sections of the Code as equal in probability of both citing and being cited.  In reality, each section contains a varying amount of language.  Thus, a more appropriate model is to assume that the probabilities of both citing and being cited are proportional to the amount of language contained in each section.  This assumption is equivalent to the following functional form:
\begin{align}
	p(\frac{\delta}{L}) \propto& (\frac{\delta}{L})^{-\alpha}
\end{align}
where $\delta$ and $L$ are the degree and amount of language of a section.  Given that the number of words per section can vary by orders of magnitude within the Code, this seems like a much more plausible model.   

In order to evaluate this model, we need to define what the ``amount of language'' $L$ means.  The most natural unit of analysis here is a single ``token,'' as each token typically corresponds to a single word.  In order to count these tokens in English text, we tokenize each section according to the Penn Treebank and let $L$ be the number of resulting strings (\cite{Marcus1994}).  This allows us to calculate $\frac{\delta}{L}$ for each section of the Code.

\begin{figure}[htb]
	\centering
	\includegraphics[width=7.5cm]{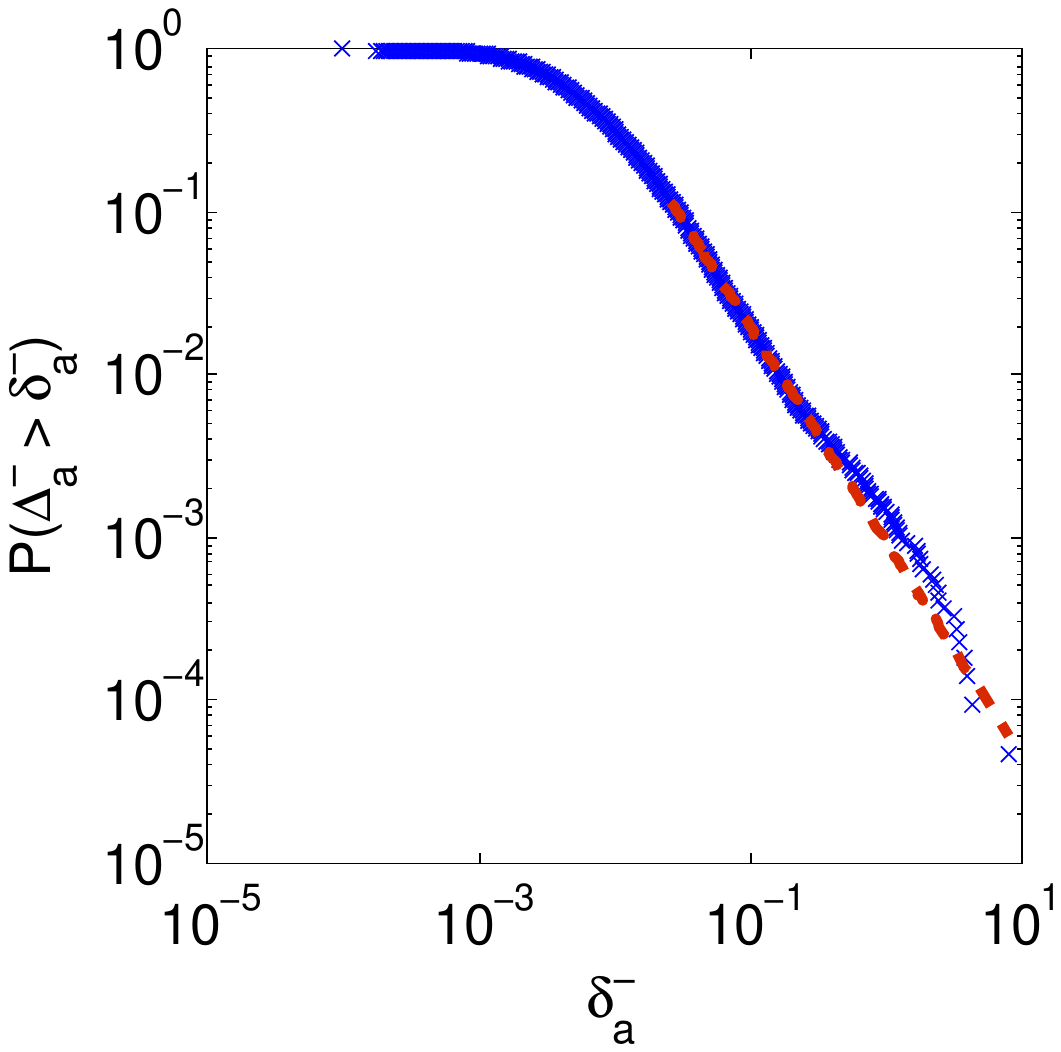}
	\caption{Log-Log Adjusted Indegree Distribution with Power-Law Fit.}
	\label{fig:adjindegree}
	
	\includegraphics[width=7.5cm]{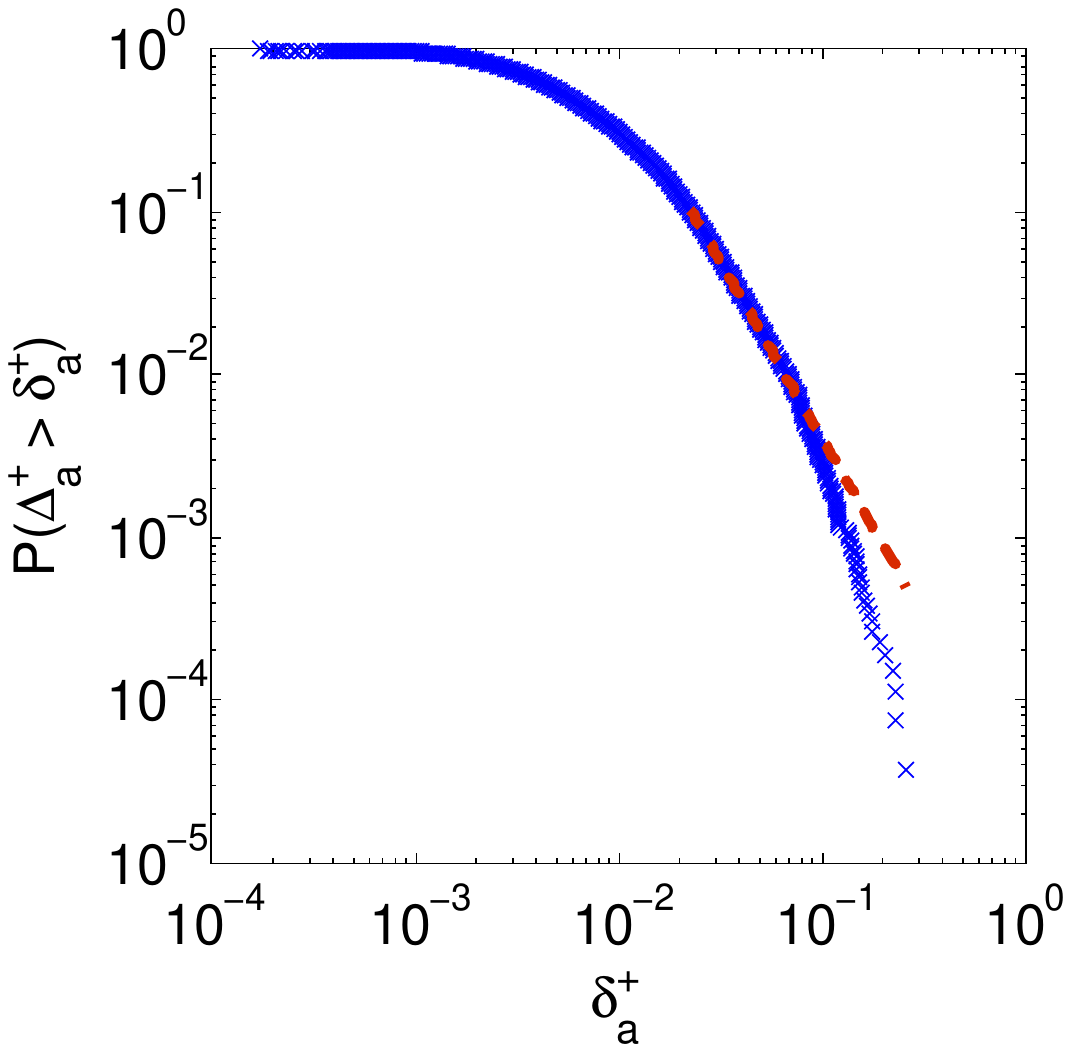}
	\caption{Log-Log Adjusted Outdegree Distribution with Power-Law Fit.}
	\label{fig:adjoutdegree}
\end{figure}

Figure \ref{fig:adjindegree} shows the log-log adjusted indegree distribution and its power-law fit.  $\alpha$ and $x_{min}$ are $2.3$ and $0.026$ respectively.  Though $\alpha$ lies between 2 and 3, the $p$-value is 0.003, soundly rejecting the model of a power-law generating process.  This rejection is due to the significant convexity displayed for small values of the distribution.

Figure \ref{fig:adjoutdegree} plots the log-log adjusted outdegree distribution and its power-law fit.  $\alpha$ and $x_{min}$ are $3.2$ and $0.023$ respectively.  In this case, $p \approx 0$ and thus the power-law model is once again incredibly implausible.  In this case, the convexity of the log-log distribution is even more pronounced.

\begin{figure}[htb]
	\centering
	\includegraphics[width=7.5cm]{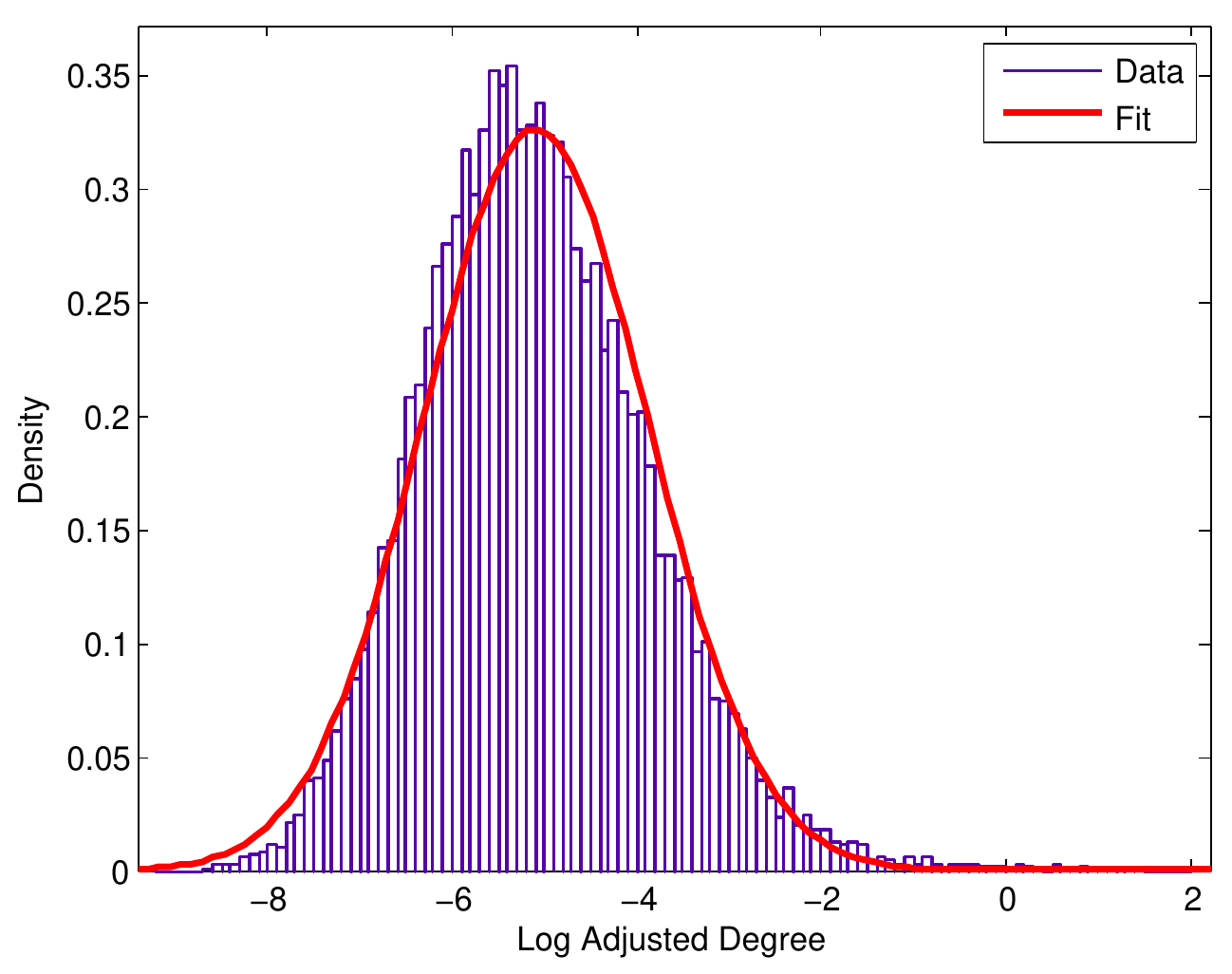}
	\caption{Log Adjusted Indegree Distribution with $\mathcal{N}(-5.09, 1.49)$ Fit.}
	\label{fig:histadjindegree}
	
	\includegraphics[width=7.5cm]{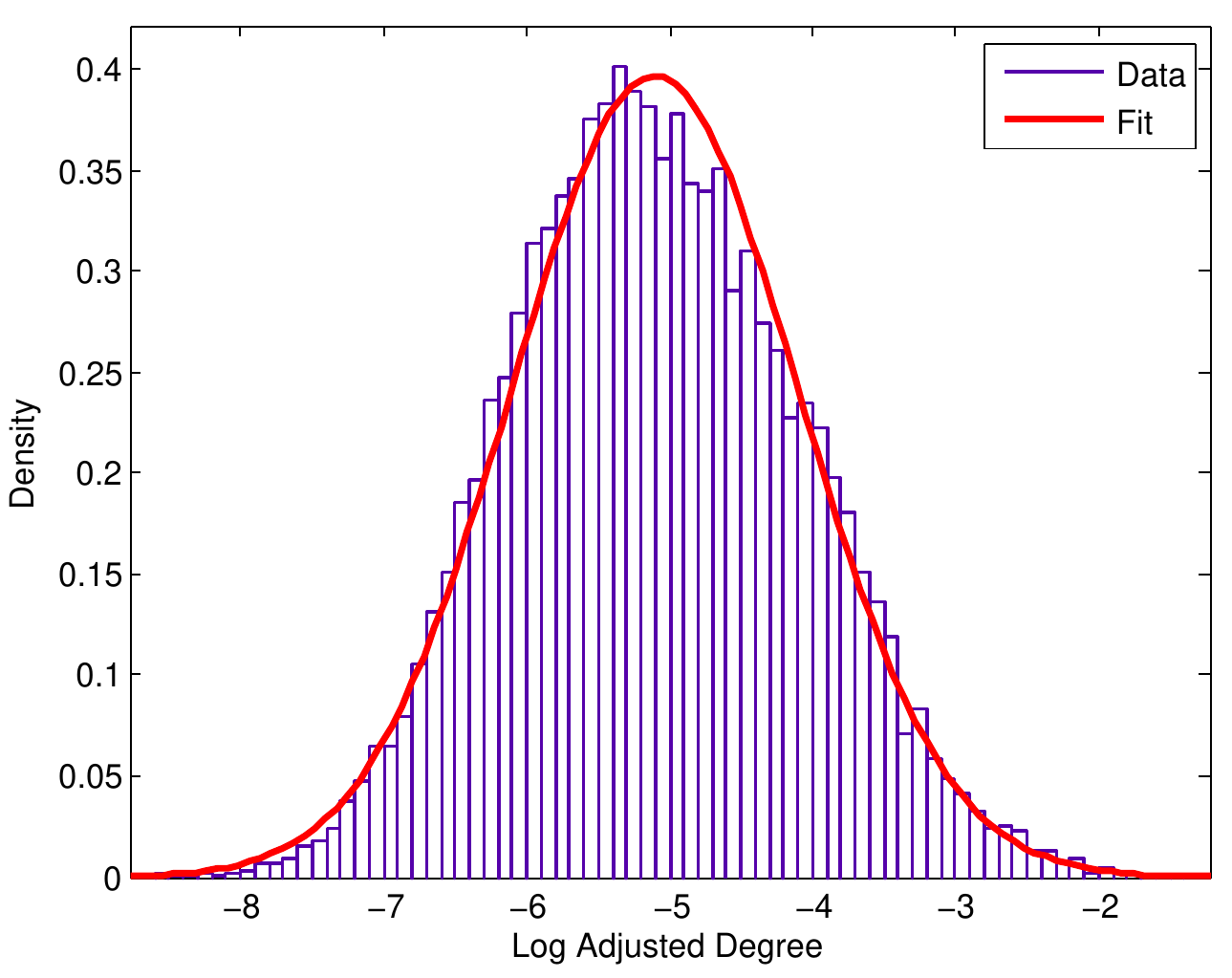}
	\caption{Log Adjusted Outdegree Distribution with $\mathcal{N}(-5.10, 1.01)$ Fit.}
	\label{fig:histadjoutdegree}
\end{figure}

These calculations indicate that the adjusted degree distributions do not seem to follow a power-law form.  They do, however, appear to approximately obey a log-normal distribution of the form:
\begin{align}
	p(\frac{\delta}{L}) \propto& \frac{L}{\delta \sigma} \exp(-\frac{(\ln(\frac{\delta}{L}) - \mu)^2}{\sigma^2})
\end{align}
where $\mu$ and $\sigma$ are the usual mean and standard deviation.  

Figure \ref{fig:histadjindegree} displays the distribution of log adjusted indegree and the corresponding normal fit with $\mu = -5.09$ and $\sigma = 1.49$.  These estimates are significant at the $p=0.05$ level and have standard errors on the order of $10^{-3}$.  The figure indicates that this distribution is a good fit, though the slight positive skewness results in asymmetric tails.  This confirms our model that the logarithm of the rate of citations received per word is normally distributed.

Figure \ref{fig:histadjoutdegree} plots the distribution of log adjusted outdegree and the corresponding normal fit with $\mu = -5.10$ and $\sigma = 1.01$.  These estimates are significant at the $p=0.05$ level and have standard errors on the order of $10^{-3}$.  The figure shows that this distribution is an excellent fit with nearly symmetric tails.  This confirms our model that the logarithm of the rate of citations made per word is normally distributed.  These adjusted degree results also imply that the model of log-normal rate of citation per word is a good fit to the data.  This is an interesting result that contradicts the conclusions of citation analyses in academic journals (\cite{Peters1994}). 

\section{Conclusion}
\label{sec:conclusion}
In this paper, we study the section citation network of the United States Code, the output of a complex and dynamic legislative process.  We find that a power-law model is not a good fit for the degree distributions, though it is somewhat plausible for the outdegree distribution.  This result indicates that the common generative mechanisms for the power-law seen in other settings are not applicable.

In order to better understand the citation dynamics in the Code, we evaluate an alternative hypothesis that the rate of citation is proportional to the amount of language within each section.  Under this adjusted model, we find that the power-law is soundly rejected for both indegree and outdegree.  Instead, the log-normal distribution is an excellent fit for both indegree and outdegree.  This implies the unintuitive result that the logarithm of the number of citations per word follows a normal distribution.  In summary, our findings indicate that the power-law is not generally applicable to citations in the United States Code but that the rate of citations per word obeys a log-normal distribution.

\section{Acknowledgments}
We would like to thank the Center for the Study of Complex Systems (CSCS) at the University of Michigan for a fruitful research environment.  This work was partially supported by an NSF IGERT fellowship through the Center for the Study of Complex Systems (CSCS) at the University of Michigan, Ann Arbor.  We would also like to thank Dave Shetland at the Legal Information Institute at the Cornell University Law School as well as Brian Karrer, Joel Slemrod, J.J. Prescott, and Abe Gong at the University of Michigan for their assistance and feedback.

\end{document}